\documentclass[a4,twocolumn,traditabstract]{aa}
\usepackage{color}
\usepackage{amsmath,amssymb,graphics,graphicx,caption,verbatim}
\usepackage{rotating}
\usepackage{subfigure}

\definecolor{eknotes}{rgb}{0.62, 0.0, 0.77}

\usepackage[varg]{txfonts}

\begin{document}

\titlerunning{Spatially inhomogeneous acceleration}
\authorrunning{Stackhouse \& Kontar}

\title{Spatially inhomogeneous acceleration of electrons in solar flares}

\author{$^{1,2}$ and $^{1}$}

\author{Duncan J. Stackhouse
          \inst{1,2}
          \and
          Eduard P. Kontar \inst{1}
          }

   \institute{School of Physics and Astronomy, University of Glasgow,
              Kelvin Building, Glasgow G12 8QQ, UK\\
         \and
             King's College London, Strand, London, WC2R 2LS, UK\\
             \email{duncan.stackhouse@kcl.ac.uk}
             }

\offprints{D.J. Stackhouse \email{duncan.stackhouse@kcl.ac.uk} }

\date{Received ; Accepted}

 \abstract{The imaging spectroscopy capabilities of the Reuven Ramaty high energy
  solar spectroscopic imager (RHESSI) enable the examination of the accelerated
  electron distribution throughout a solar flare region.
  In particular, it has been revealed that the energisation of these particles takes place   over a region of finite size, sometimes resolved by RHESSI observations.
  In this paper, we present, for the first time, a spatially distributed acceleration model and investigate the role of inhomogeneous acceleration on the observed X-ray emission properties. We have modelled transport explicitly examining scatter-free and diffusive transport within the acceleration region and compare with the analytic leaky-box solution. The results show the importance of including this spatial variation when modelling electron acceleration in solar flares. The presence of an inhomogeneous, extended acceleration region produces a spectral index that is, in most cases, different from the simple leaky-box prediction.
  In particular, it results in a generally softer spectral index than predicted by the leaky-box solution, for both scatter-free
  and diffusive transport, and thus should be taken into account when modelling
  stochastic acceleration in solar flares.}


  \maketitle

 \section{Introduction}\label{intro}

  Large solar flares can release up to $\sim 10^{32}$~erg of
  energy due to the restructuring of the Sun's magnetic field \citep[e.g.][]{2012ApJ...759...71E}. Flares
  have long been known
  to accelerate particles in the corona \citep{1959JGR....64..697P},
  but the processes behind the transfer of this magnetic energy
  from reconnection is not fully understood \citep{2011SSRv..159..107H}.
  The Reuven Ramaty high energy
  solar spectroscopic imager (RHESSI) \citep{2002SoPh..210....3L} enabled, for the first time,
  observations of the deka-keV hard X-ray (HXR) spectrum well resolved
  in energy, space and time \citep[see][for recent reviews]{2011SSRv..159..107H,2011SSRv..159..301K}.
  The imaging spectroscopy capabilities of RHESSI allows new avenues of investigation;
  \citet{2003ApJ...595L.107E}, \citet{2006A&A...456..751B} and \citet{2010ApJ...712L.131P}
  use spatially resolved
  images of looptop and footpoint sources to compare the electron spectrum
  throughout the HXR source whereas \citet{2005ApJ...629L.137L}, \citet{2009ApJ...696..121L} and
  \citet{2013ApJ...766...75J}
  investigate the time dependence of the shape of the looptop sources.
  Of particular note is the resolution of the acceleration region, showing that to be
  consistent with observations it must be extended in space \citep[e.g.][]{2008ApJ...673..576X,2011ApJ...730L..22K,2012A&A...543A..53G}.

  Acceleration in the coronal plasma can be split into two broad
  regimes, whether the process behind it is systematic or stochastic
  in nature. Observational evidence \citep{2006ApJ...653L.149K} points
  toward an accelerated electron population that is isotropic, favouring
  a stochastic acceleration mechanism. Furthermore, systematic acceleration regimes
  often have large scale electrodynamic issues intrinsic within them
  \citep{1995ApJ...446..371E}. Stochastic acceleration, also called
  second order Fermi acceleration \citep{1949PhRv...75.1169F},
  also produces acceleration efficiencies consistent
  with HXR observations \citep{2008AIPC.1039....3E}. The actual process
  of stochastically accelerating electrons can happen in a variety of
  ways \citep{2012ApJ...754..103B} but the acceleration itself is
  most often well described by a turbulent diffusion coefficient, $D_{vv}$ \citep{1966PhRv..141..186S,1968Ap&SS...2..171M}.

  As the particles are accelerated in the corona they move through it, some
  reaching lower levels of the solar atmosphere. In most cases this results
  in electrons at deka-keV energies reaching the chromosphere where they emit
  as HXR footpoints \citep{1986SSRv...44...43D,1988psf..book.....T,2011SSRv..159..107H}, with the non-thermal looptop
  spectrum being relatively softer \citep{2006A&A...456..751B}. If the
  density is high enough within the accelerating region there will be cases where the
  HXR emission is confined to the corona \citep{2008ApJ...673..576X}, this being
  the subject of our study in \citet{2014ApJ...796..142B}. The first
  RHESSI observations of coronal thick targets are described
  by \citet{2004ApJ...603L.117V}.

  In an X-ray context, the
  photon spectrum from the looptop has a thermal-like core and a power-law, or broken
  power-law tail. The footpoint spectrum also has a thermal component, likely with
  a lower temperature than the looptop source, with a non-thermal tail having a relatively
  harder spectral index than the coronal spectrum \citep{2003ApJ...595L.107E, 2006A&A...456..751B}.
  The electron spectrum producing this photon spectrum can be inferred by a variety of
  techniques such as: forward-fitting \citep{2003ApJ...595L..97H}, regularised  inversion
  \citep{2003ApJ...595L.127P, 2004SoPh..225..293K}, or inversion with data-adaptive
  binning \citep{1992SoPh..137..121J}. The strengths and weaknesses of these methods for
  reproducing features present in the electron spectrum are discussed in
  \citet{2006ApJ...643..523B}.

  Transport of electrons of tens of keV in solar flares could be expected to fall into
  one of two categories, scatter-free (no pitch-angle scattering) or diffusive
  (pitch-angle scattering). If the transport is scatter-free in nature the
  accelerated electrons experience negligible pitch-angle scattering
  and hence, for sufficiently high velocities, deposit most of their
  energy in the dense chromospheric footpoints.
  There is mounting evidence, however, that the electrons
  should be scattered: firstly, there is a lack of anisotropy evident
  from hard X-ray observations \cite[][as a review]{2011SSRv..159..301K};
  secondly, albedo diagnostics \citep{2006ApJ...653L.149K,2013SoPh..284..405D} as well
  as stereoscopic measurements \citep{1998ApJ...500.1003K} are inconsistent
  with strong downward beaming below $\sim 100$~keV; thirdly, the majority of stochastic acceleration
  models developed for solar flares require strong pitch-angle scattering
  \citep{1966PhRv..141..186S,1968Ap&SS...2..171M,1987SoPh..107..299B,1999ApJ...527..945P};
  finally, the accelerated electrons will propagate in a turbulent or beam-generated
  turbulent media.

  Interestingly, the advent of RHESSI imaging spectroscopy
  \citep{2002SoPh..210...61H} confirmed earlier work using the
  Yohkoh spacecraft \citep[e.g.][]{2002ApJ...569..459P}
  that the photon spectral index difference between looptop and footpoint sources was not two as would be expected in the thick-target model
  \citep{2003ApJ...595L.107E,2006A&A...456..751B,2008SoPh..250...53S,2010ApJ...712L.131P}.
  Furthermore, \citet{2014ApJ...780..176K}
  show that the electron injection rates at the looptop
   are more than is required to produce the footpoint emission. Introducing an effective mean free path, $\lambda$,
  parallel to the magnetic field to account for the effect of pitch-angle diffusion of
  particles they find that this should be $10^8-10^9$~cm,
  which is less than the length of a loop and comparable to the size
  of the acceleration region.

  As already mentioned, RHESSI imaging spectroscopy has revealed that the
  acceleration region in the hard X-ray looptop sources occupy a noticeable fraction
  of the loop \citep{2008ApJ...673..576X,2011ApJ...730L..22K}.
  So far, however, the modelling and comparison with observations has been
  limited to spatially averaged or single-point acceleration or injection.
  Current models, for example the leaky-box approximation, account for transport implicitly
  by introducing an escape term. This allows
  the study of the acceleration term without complications arising from transport \citep[e.g.][]{2013ApJ...777...33C}.
  While the energy distribution can be studied, the spatial distribution
  observed in flares cannot.
  An alternative simplifying approach is to inject an
  already accelerated power-law electron distribution and examine various transport effects
  \citep[e.g.][]{1982ApJ...259..341B,1983SoPh...86..133E,1990ApJ...359..524M,1991ApJ...368..316R,2014ApJ...787...86J},
  but this does not account for the effects of acceleration
  on the transport process. Evidently, such a split between acceleration and
  transport is not justified and inadequate to model recent RHESSI observations.

  In this paper, we develop a model that accounts simultaneously for the
  transport and acceleration of electrons in a model of the solar corona. We examine the effects of a spatially varying, extended
  acceleration region and determine how the electron spectrum evolves from
  an initial Maxwellian distribution. We find that the introduction of an
  extended, inhomogeneous, acceleration region results in a spectrum that is, in general, softer for both scatter-free
  and diffusive transport when comparing
  to the spectral index expected from the analytic leaky-box solution. The authors therefore suggest that explicit
  spatial effects should be taken account of when modelling acceleration and transport in
  solar flares.

  Section \ref{model_sect} introduces the model describing  the acceleration and parallel
  transport in solar flares.  Section \ref{RHESSI_sect} discusses the derived parameters
  and summarises the observational results. In Section \ref{numerical_results} we show the
  results of our numerical simulations comparing them to the leaky-box solution and
  to the imaging spectroscopy results for context.  Section \ref{conc_sect} discusses
  the implications and possibilities of further work.

 \section{Acceleration and transport of energetic electrons in solar flares}\label{model_sect}

    The evolution of the electron phase space distribution, $f$,
    parallel to the magnetic field, $\mathbf{B}_0$
    (aligned in the x-direction), can be described by the Fokker-Planck equation. In the next two sub-sections we outline the two transport regimes studied in the paper.

    \subsection{Scatter-free transport}

    If the electron accelerating current is field aligned, that is parallel to the background magnetic field, then
    the electron dynamics can be approximated as one-dimensional in velocity. In this case stochastic acceleration only
    acts to accelerate electrons parallel to the field and so
    the evolution of the electron phase space distribution, $f(v,x,t)$ [e$^{-}$ cm$^{-4}$ s], is described by the
    one-dimensional Fokker-Planck equation,
    \begin{equation}\label{1dfp}
    \frac{\partial f}{\partial t} + v \frac{\partial f}{\partial x}  = \frac{\partial}{\partial v}\left[D(v,x)
                                                                      + \frac{\Gamma(x) v_{\rm te}^2}{v^3}\right]
                                                                      \frac{\partial f}{\partial v}
                                                                      + \Gamma(x) \frac{\partial}{\partial v}\left(\frac{f}{v^2}\right),
    \end{equation}
    where $v_{\rm te}=\sqrt{k_{\rm B} T /m_{\rm e}}$~[cm s$^{-1}$]
    is the thermal speed, with temperature, $T$~[K], $k_{\rm B}$~[erg K$^{-1}$] boltzmann's
    constant and $m_{\rm e}$~[g] the mass of an electron.
    The collisional parameter is $\Gamma = 4 \pi e^4 \ln \Lambda n(x) / m_{\rm e}^2$~[cm$^3$ s$^{-4}$], with
    $e$ [e.s.u] the electron charge, $n(x)$ [cm$^{-3}$] the density and $\ln \Lambda$ the coulomb logarithm
    taken to be $\simeq 20$ for solar flare conditions.
    $x$~[cm] is the distance from the top of the loop and $v$~[cm s$^{-1}$] is the velocity.
    The distribution is normalised so $n_e=\int f dv$, where $n_e$~[cm$^{-3}$] is the electron number
    density. The second term on the left hand side of Equation \eqref{1dfp} describes
    the scatter-free transport in the system,
    while the second term inside the brackets on the right is the diffusion due to
    collisions and the final term on the right hand side describes the energy loss
    due to Coulomb collisions. $D(v,x)$ [cm$^2$ s$^{-3}$] is the spatially one-dimensional turbulent diffusion
    coefficient discussed in Section \ref{dturb_sect}. We note that we use a simplified version of the collisional operator here, where the electrons are modelled as being in contact
    with a heat-bath of constant temperature, $T$. This is applicable to the
    solar flare situation, as discussed by \citet{2014ApJ...787...86J}.

    \subsection{Diffusive transport}

    In the case of strong pitch-angle ($\mu = \cos \theta$) scattering, where the mean free path due to scattering
    is less than the characteristic acceleration region length, we use the angle averaged
    three-dimensional form of the Fokker-Planck equation assuming the distribution is isotropic in
    pitch-angle. The evolution of the electron
    phase space distribution, $f(\mathbf{v},x,t)$ [e$^{-}$ cm$^{-6}$ s$^3$],
    is then,
    \begin{equation}\label{3dfp}
    \frac{\partial f}{\partial t} + \mu v \frac{\partial f}{\partial x}   = \frac{1}{v^2}\frac{\partial}{\partial v}\left[v^2 D(v,x)+ \frac{\Gamma(x) v_{\rm te}^2}{v}\right]
                                                                              \frac{\partial f}{\partial v}
                                                                              + \frac{\Gamma(x)}{v^2} \frac{\partial f}{\partial v},
    \end{equation}
    where the terms are analogous to those in Equation \eqref{1dfp},
    but for the isotropic distribution the normalisation is $n_e = \int f 4\pi v^2 dv$.
    In the regime of strong pitch-angle scattering pitch-angle diffusion
    leads to a fast flattening of the $\mu$ distribution function over time,
    i.e. $\partial f/\partial \mu \rightarrow 0$. So the transport becomes a
    spatial diffusion parallel to the magnetic field \citep{1966ApJ...146..480J, 2014ApJ...780..176K}:
    \begin{equation}
        \mu v \frac{\partial f}{\partial x} \rightarrow - D_{\rm xx} \frac{\partial^2 f}{\partial x^2} .
    \end{equation}
    We introduce a spatial diffusion coefficient, $D_{\rm xx} =\lambda(v) v/3$, where
    $\lambda$ [cm] is the mean free path accounting for non-collisional pitch-angle scattering. The expression for
    this mean free path is given in \cite{2014ApJ...780..176K}:
    \begin{equation}\label{genmfp}
      \lambda(v) = \frac{3v}{8} \int_{-1}^{1} \frac{(1-\mu^2)^2}{D_{\mu\mu}^{(T)}+D_{\mu\mu}^{(C)}} d \mu ,
    \end{equation}
    where $D_{\mu\mu}^{(C)}$ is the collisional pitch-angle diffusion coefficient
    and $D_{\mu\mu}^{(T)}$ is the turbulent pitch-angle diffusion coefficient.
    The velocity (or energy) dependence of $D_{\mu\mu}^{(T)}$  is poorly known
    in solar flares and, in principle, $\lambda(v)$ could have a complicated dependence on energy.
    In this paper, we examine one case, a constant mean free path for all velocities
    with a value $\lambda = 5 \times 10^8$~cm, as this is
    the midpoint of the limits \citet{2014ApJ...780..176K} find for $30$~keV electrons.
    Equation \eqref{genmfp} can be re-written in terms of a scattering timescale, $\tau$, so,
    \begin{equation}\label{mfpapprox}
      \lambda(v) = \frac{3v}{8} \tau (v).
    \end{equation}
    To obtain an order of magnitude estimate for the scattering timescale,     the limits from \citet{2014ApJ...780..176K} are used. Setting
    $\lambda = 5 \times 10^8$~cm at $30$~keV this gives a scattering
    timescale of $\tau(30~{\rm keV}) \simeq 0.18$~s. The results of the numerical simulations with constant mean free path are shown in Section \ref{numerical_results} together with     those from the scatter-free simulations.

    \subsection{Spatially dependent diffusion coefficient}\label{dturb_sect}

\begin{figure}[htpb]
    \centering
     \includegraphics[width=0.45\textwidth, clip=true, trim=1.5cm 1.5cm 0cm 1.5cm]{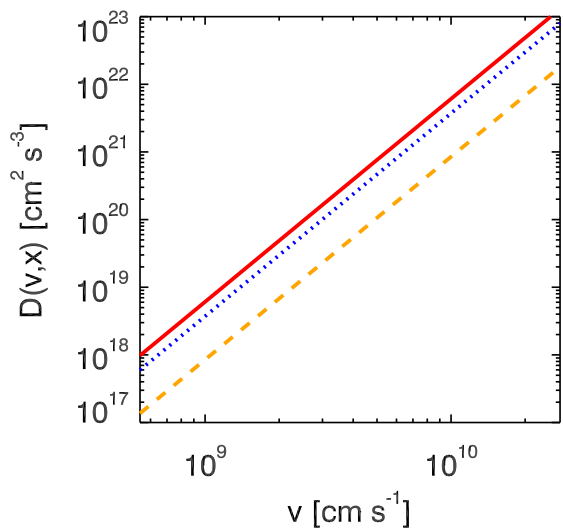}
        \caption{Diffusion coefficient versus velocity for $\tau_{\rm acc} = 10\tau^{\rm th}_{\rm c}$ for three different points
        in space: \emph{Red, solid line}; $D(v, x=0)$,
        \emph{Blue, dot line}; $D(v, x = \sigma)$ and \emph{Orange, dash line}; $D(v, x=2\sigma)$.}
        \label{diffusion_coefficient_fig}
    \end{figure}

    Imaging spectroscopy with RHESSI has revealed the extended nature of the acceleration region in the HXR looptop source \cite[e.g.][]{2008ApJ...673..576X,2012A&A...543A..53G}.
    In order to examine the effects of a spatially dependent, extended
    acceleration region in a regime with simultaneous transport,
    we introduce a spatially non-uniform velocity diffusion coefficient:
     \begin{equation}\label{dturb}
     D(v,x) = \frac{v_{\rm te}^2}{\tau_{\rm acc}}\left(\frac{v}{v_{\rm te}}\right)^{\alpha}
     \operatorname{e}^{-x^2/2 \sigma^2},
    \end{equation}
    where $\tau_{\rm acc}$ [s] is the acceleration timescale, $\sigma$ [cm] is the
    spatial extent of the acceleration region and $\alpha$ is a constant that
    controls the strength of the velocity dependance.
    With this choice, we confine the acceleration to a region
    in space, akin to an extended looptop acceleration region,
    as observed by RHESSI \cite[e.g.][]{2008ApJ...673..576X,2011ApJ...730L..22K,2012A&A...543A..53G}.
    Equation \ref{dturb} assumes that the acceleration efficiency
    within this region is most effective at $x=0$, the top of the loop,
    and that there is a drop off with distance that is Gaussian in nature.
    The length of the acceleration region,
    density and temperature are determined from RHESSI imaging spectroscopy
    and  discussed in Section \ref{RHESSI_sect}.
    The diffusion coefficient (Equation \ref{dturb}) is shown in Figure \ref{diffusion_coefficient_fig}
    for a specific choice of acceleration timescale, $\tau_{\rm acc}$, spatial extent, $\sigma$, and thermal
    velocity, $v_{\rm te}$, the latter two obtained from imaging spectroscopy (see Section \ref{RHESSI_results_sect}).

   \subsection{The leaky-box Fokker-Planck approximation}\label{leaky_sect}

    At this point it is instructive to examine the leaky-box Fokker-Planck
    approximation \cite[e.g.][]{1969ApJ...156..445K,1977ApJ...211..270B,2013ApJ...777...33C}.
    This model is a spatially averaged description of the velocity (or energy) evolution of
    electrons in the acceleration region designed to study the spectral properties of accelerated electrons, and has been used as a comparison to solar flares.
    Replacing transport by an escape timescale term we have the equation for the
    spatially-averaged distribution function, $\langle f(v,t)\rangle$,
    \begin{equation}\label{leaky}
         \frac{\partial \langle f \rangle}{\partial t} =  \frac{\partial}{\partial v}\langle D(v)\rangle\frac{\partial \langle f\rangle}{\partial v}
         - \frac{\langle f\rangle}{\tau_{\rm esc}(v)},
    \end{equation}
    where,
    \begin{align}
    \langle D(v)\rangle & = \frac{v_{\rm te}^2} {\tau_{\rm acc} } (v/v_{\rm te})^{\alpha}
    \sim (2 \pi L)^{-1/2} \int_{-L/2}^{L/2} D(v,x),\notag \\ & =
    \frac{v_{\rm te}^2} {\tau_{\rm acc}}  (v/v_{\rm te})^{\alpha}
    (2 \pi L)^{-1/2} \int_{-L/2}^{L/2} \exp(-x^2/2 \sigma^2),
    \end{align}
    is the acceleration region averaged velocity diffusion coefficient
    and $\langle ... \rangle$ denotes spatial averaging over the full width half maximum (FWHM), $L=2.35\sigma$.
    Equation (\ref{leaky}) is informative and simple to use, but
    ignores the essential
    spatial dependencies in acceleration and transport. The stationary
    solution
    as $t \rightarrow \infty$ can be readily obtained from the following
    equation,
    \begin{equation}\label{eq:stat_leakybox}
         0 =  \frac{\partial}{\partial v}\langle D(v)\rangle \frac{\partial \langle f \rangle}{\partial v}
         - \frac{ \langle f \rangle}{\tau_{\rm esc}(v)}.
    \end{equation}
    Since the X-ray producing electron spectrum can often be approximated by a power-law
    \citep{2003ApJ...595L..97H}, a stationary solution of Equation (\ref{eq:stat_leakybox})
    in the form $\langle f\rangle \sim v^{-\delta_1}$ is assumed.
    Substituting this power-law solution of $\langle f\rangle$, we have,
    \begin{equation}
      \frac{\partial}{\partial v} \frac{v_{\rm te}^2}{\tau_{\rm acc}}\left(\frac{v}{v_{\rm te}}\right)^{\alpha} \frac{\partial v^{-\delta_1}}{\partial v} - \frac{ v^{-\delta_1}}{\tau_{\rm esc}(v)} = 0.
    \end{equation}
    Differentiating this expression and rearranging we
    can find an expression for $\delta_1$, the power-law index.
    For scatter-free transport, the escape timescale is equal
    to the free streaming timescale, $\tau_{\rm esc} = \sigma/v$. Therefore, the spectral index is,
   \begin{equation}
     \delta_1 = \frac{1}{2}\left[\alpha -1 + \left( (1-\alpha)^2 +4  \frac{\tau_{\rm acc}v_{\rm te}^{\alpha-2}}{\sigma} v^{3-\alpha}   \right)^{1/2}    \right],
   \end{equation}
   where one sees that a power-law electron spectrum ($v$ independent $\delta _1$) can be obtained only for $\alpha =3$, so,
   \begin{equation}\label{leaky_scatter-free_delta}
     \delta_1(\tau_{\rm acc}) = \frac{1}{2}\left[2 + \left( 4 +4 \frac{v_{\rm te}}{\sigma} \tau_{\rm acc} \right)^{1/2}  \right].
   \end{equation}
   Of course, in order to put our results here, and those of the numerical simulations, in the
   context of the imaging spectroscopy results of Section \ref{RHESSI_sect} we need the index
   of the density weighted mean electron flux $\langle n V F(E) \rangle$. Using the fact that
   $\langle n V F(E) \rangle \sim \langle f \rangle / m_{\rm e}$ in one-dimension this means that,
   \begin{equation}\label{scatt_free_LT_E_spect}
      \langle n V F(E) \rangle^{\rm 1d}_{\rm LT} \sim v^{-\delta_1} \sim E^{-\delta_1/2},
   \end{equation}
   where the superscript makes clear this is the one-dimensional scatter-free expression and the subscript shows that
   this is the expected $\langle n V F(E) \rangle$ from the looptop.

   Similarly, the three-dimensional Fokker-Planck (Equation \ref{3dfp}) gives
   the power-law index,
    \begin{equation}
     \delta_2 = \frac{1}{2}\left[\alpha +1 + \left( (\alpha+1)^2 + 4  \frac{\tau_{\rm acc}}{\tau_{\rm esc}} \left(\frac{v_{\rm te}}{v}\right)^{\alpha-2}   \right)^{1/2}    \right],
    \end{equation}
    where $\tau_{\rm esc} = 3 \sigma^2/\lambda(v) v$ \citep[e.g.][]{2014ApJ...796..142B} and
    $\lambda$ is the mean free path of an electron due to pitch-angle scattering. For constant $\lambda$, the power-law $\langle f\rangle $ again requires $\alpha =3$, so,
    \begin{equation}\label{leaky_diffusive_delta}
     \delta_2(\tau_{\rm acc}) = \frac{1}{2}\left[4 + \left( 16 + 4  \frac{\lambda(v) v_{\rm te}}{3 \sigma^2} \tau_{\rm acc} \right)^{1/2}    \right].
    \end{equation}
    As before, but for the three-dimensional Fokker-Planck, $\langle n V F(E) \rangle \sim v^2 f(v)/ m_{\rm e}$, so
    we have,
    \begin{equation}\label{isotropic_LT_E_spect}
        \langle n V F(E) \rangle_{\rm LT}^{\rm 3d} \sim E f(v) = E v^{\delta_2} \sim E^{-\delta_2/2 + 1},
    \end{equation}
    where the superscript and subscript illustrates that this is the three-dimensional Fokker-Planck with diffusive transport for the looptop spectrum.

    So, the above arguments give us the looptop spectral index predicted by the leaky-box
    Fokker-Planck solution, $\langle n V F(E) \rangle^{\rm 1d}_{\rm LT} \sim E^{-\delta_1/2}$ or
    $\langle n V F(E) \rangle^{\rm 3d}_{\rm LT} \sim E^{-\delta_2/2+1}$,
    depending on whether there is negligible or strong pitch-angle scattering respectively. In order to
    find the footpoint spectrum predicted in both cases, one needs the electron
    escape rate from the looptop source, $\dot{N} (E)$
    [e$^{-}$ s$^{-1}$ per unit energy].
    The number of particles per second per unit speed, $\dot{N}(v)$ [e$^{-}$ s$^{-1}$ (cm s$^{-1}$)$^{-1}$], is the flux multiplied by the volume, that is,
    \begin{equation}
      \dot{N}(v) = \frac{\langle f \rangle_{\rm LT}}{\tau_{\rm esc}} V .
    \end{equation}
    Now, since for the one-dimensional Fokker-Planck the total number is $n_e = \int f dv$ this means that
    $\dot{N}(E) ~ dE = \dot{N}(v) ~ dv$ and so,
    \begin{equation}\label{dotN_eqn_1D}
      \dot{N}^{\rm 1d}(E) = \frac{1}{m_e v} \frac{\langle f \rangle_{LT}}{\tau_{\rm esc}} V,
    \end{equation}
    the total number for the three-dimensional case, however, is $n_e = \int f 4 \pi v^2 dv$, so $\dot{N}(E) ~ dE = \dot{N}(v) ~ 4 \pi v^2 dv$
    and,
    \begin{equation}\label{dotN_eqn_3D}
      \dot{N}^{\rm 3d}(E) = \frac{4 \pi v}{m_e} \frac{\langle f \rangle_{LT} }{\tau_{\rm esc}} V.
    \end{equation}
    The density weighted mean electron flux at the footpoint is given by,
    \begin{equation}
      \langle n V F(E) \rangle_{\rm FP} = \frac{E}{K}  \int_E^{\infty} \dot{N}(E) dE,
    \end{equation}
    and so for the one-dimensional case this gives,
    \begin{equation}
      \langle n V F(E) \rangle^{\rm 1d}_{\rm FP} = \frac{V}{m_e K \sigma}  E \int_E^{\infty} \frac{1}{v} \langle f \rangle_{LT} v dE,
    \end{equation}
    which means,
    \begin{equation}\label{scattfree_FP_E_spect}
      \langle n V F(E) \rangle^{\rm 1d}_{\rm FP} \propto E \int_E^{\infty} E^{-\delta_1/2} dE \sim E^{-\delta_1/2+2} .
    \end{equation}
    A similar argument leads to,
    \begin{equation}\label{isotropic_FP_E_spect}
      \langle n V F(E) \rangle^{\rm 3d}_{\rm FP} \propto E^{-\delta_2/2 + 3},
    \end{equation}
    for the three-dimensional case.

    In both cases, the power-law spectral index depends on the value
    of $\tau_{\rm acc}$. If there is point like acceleration at the apex of the
    loop, with this configuration, one might expect a spectral index close to
    $\delta_1$ or $\delta_2$ to form.  However, the spatial non-uniformity of the
    acceleration region results in local acceleration times given by,
    \begin{equation}\label{effective_timescale}
      \tau_{\rm eff}(x) = \tau_{\rm acc} \exp \left(\frac{x^2}{ 2 \sigma^2}\right),
    \end{equation}
    due to $x$ dependency of $D(v,x)$
    (equations \ref{1dfp} and \ref{3dfp}) and hence a different local distribution function.
    Therefore, a spatially dependent acceleration region creates different spectral
    indices at each point in space. The resulting distribution function from the
    entire acceleration region is controlled by the transport between various spatial locations.
    The resulting spectral index (if a power-law is formed) could be different from that
    predicted by our leaky-box solution.

    Table \ref{spectra_table} shows the relationship between the spectral indices of the electron phase-space distribution,
    density weighted mean electron flux and photon spectrum. When comparing models in Section \ref{numerical_results} we use the
    spectral indices of the density weighted mean electron flux, for the reasons discussed in \citet{2003ApJ...595L.115B}.

    These results are compared to numerical simulations with non-spatially averaged acceleration and transport.
    The importance of including the spatial dependence is shown clearly in Sections \ref{scatter-free_sect} and
    \ref{diffusive_sect}.

  \begin{table*}
    \begin{center}
    \caption{Summary of the pertinent spectra in this paper and the relationship of their spectral indices. (Electron-Ion bremsstrahlung
     is the dominant process below $\sim$100~keV.)}\label{spectra_table}
     \begin{tabular}{  l  l  c }
    \hline
    Symbol & Description & Spectral Index \\ \hline \hline
    $f(v)$ & electron speed distribution (one-dimensional) & $\delta_1$ \\
    $f(\mathbf{v})$ & electron velocity distribution (three-dimensional) & $\delta_2$ \\ \hline
    $\langle nVF(E)\rangle^{\rm 1d}$ & density weighted mean electron flux (one-dimensional) & $\delta = \delta_1/2$ (LT) \newline $\delta = \delta_1/2$ (FP) \\
    $\langle nVF(E)\rangle^{\rm 3d}$ & density weighted mean electron flux (three-dimensional) & $\delta = \delta_2/2 + 1$ (LT) \newline $\delta = \delta_2/2 + 3$ (FP) \\ \hline
    $I(\epsilon)$ & photon spectrum & $\gamma \simeq \delta + 1$ (for electron-ion bremsstrahlung) \\
    \hline
     \end{tabular}
    \end{center}
  \end{table*}

    \section{RHESSI observations and properties of non-thermal electrons}\label{RHESSI_sect}

    We use observations from a well studied flare (24 Feb 2011
    07:29:40 - 07:32:36 UT) to derive the properties of the acceleration region which are used as the input for our model simulations. This event was chosen due to it being on the limb, thus enabling easy selection of the looptop  and footpoint sources.
    Further to this, the looptop source has enough high energy photons to adequately
    constrain the non-thermal population of electrons present there.

    Using the CLEAN algorithm \citep{2002SoPh..210...61H} with a beam width
    parameter of $1.9$ \citep{2013A&A...551A.135S}, the resulting image is shown
    in the top panel of Figure \ref{spectrum}. The regions were chosen
    to have no overlap to avoid cross-contamination
    and the photon spectra obtained from the looptop and footpoint regions were forward-fit
    \citep[see e.g.][]{2003ApJ...595L..97H} with the vth and thin2
    functions in OSPEX \citep{2002SoPh..210..165S}.
    The fits yield a density weighted mean electron flux, $\langle n V F(E) \rangle$
    [e$^-$ cm$^{-2}$ s$^{-1}$ keV$^{-1}$], suitable for comparison
    with either the thin- or thick- target model. While it may seem incongruous to fit the emission from the dense chromosphere with a function containing thin-target bremsstrahlung; in reality, since we are seeking to
    compare the observations to numerical results, it only matters that
    we assume the same bremsstrahlung cross-section in both cases. It does not matter which fit function you use for the non-thermal part
    of the photon spectrum, so long as you use the same for your numerical results. \citet{2003ApJ...595L.115B} discuss the reasoning behind using
    $\langle n V F(E)\rangle$ as the natural middle ground when comparing
    observations to numerical simulations of the HXR spectrum.

    For the non-thermal population of electrons the density weighted mean
    electron flux is \citep{2013A&A...551A.135S},
      \begin{equation}\label{nvf_nontherm}
        \langle n V F(E) \rangle_{\rm nth} = \langle n V F_0(E) \rangle \frac{ \delta-1}{E_{\rm c}} \left( \frac{E}{E_{\rm c}} \right)^{-\delta}, ~ E>E_{\rm c}\;,
      \end{equation}
    where $\langle n V F_0(E) \rangle$ [e$^-$ cm$^{-2}$ s$^{-1}$] is the normalisation
    flux obtained from the OSPEX fit, $\delta$ is the fitted
    power-law spectral index, and $E_{\rm c} = 20$~keV is kept constant for each fit. The thermal part of the spectrum provides the emission measure, $EM$ [cm$^{-3}$],
    and the temperature, $k_{\rm B} T$ [keV], in the coronal part of the loop,
    so that the density weighted mean electron flux \citep{1988ApJ...331..554B,2013ApJ...779..107B} is,
     \begin{equation}
       \langle n V F(E) \rangle_{\rm th} = EM \frac{2^{3/2}}{(\pi m_{\rm e})^{1/2}} \frac{E}{(k_{\rm B} T)^{3/2}} \operatorname{e}^{-E/k_{\rm B} T}.
     \end{equation}

    \subsection{Thermal and spatial source parameters}

    \begin{figure}[htpb]
    \centering
     \includegraphics[width=0.45\textwidth]{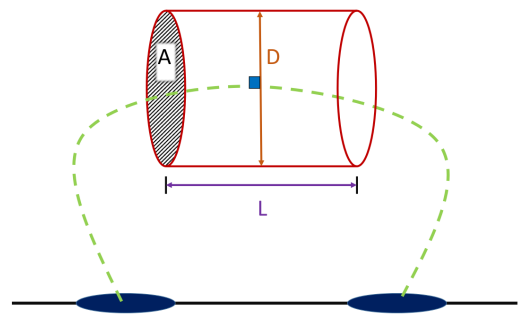}
        \caption{Sketch of the geometry showing parameters used to calculate $V_{\rm th}$, $n_e$, and $L$. The chromosphere is shown by the black solid line, with the HXR footpoints shown in blue. The loop midline is shown by the green dashed line. The diameter, $D$, is shown by the orange line. The length, $L$, is shown by the purple line and the cross-sectional area, A, is shown by the shaded end.}
        \label{loop_cartoon}
    \end{figure}

    For estimating the cross-sectional area, and thus the volume of the
    thermal source, we assumed a loop like geometry joined to the chromosphere at the footpoints.
    The loop morphology of coronal sources has been extensively discussed in general, but not for this event,
    by \citet{2008ApJ...673..576X,2011ApJ...730L..22K}.
    Figure \ref{loop_cartoon} shows a cartoon of the CLEAN image in the top panel of
    Figure \ref{spectrum} highlighting the pertinent measurements. The cross-sectional
    area is assumed to be circular, $A=\pi D^2/4$, with diameter, $D$,
    being estimated by first identifying the maximum emission in the energy band,
    $10-11.4$~keV (as we are calculating the thermal volume to estimate the
    density from the emission measure, i.e. the thermal fit), then finding the distance
    bounded by the 50\% contours and approximately orthogonal
    to the `loop midline.' The thermal volume,
    $V_{\rm th}$, is then calculated by multiplying the area, $A$,
    by the length of looptop emission, $L$, which is obtained by approximating
    the length along the loop midline and again bounded by the $50\%$ contours,
    i.e. the FWHM of the thermal emission. The spatial extent of the
    acceleration region is assumed to be the standard deviation
    of the full width half maximum as in \citet{2008ApJ...673..576X} and is given by $\sigma=L/2.35$.

    Using $EM = \bar{n}^2 V$ from thermal fit we obtained an estimate of the mean target proton density, $\bar{n}=n_{\rm protons} = n_{\rm electrons}$ assuming a Hydrogen plasma. The looptop source is best fit by an emission measure,
    $EM = (0.12 \pm 0.04) \times 10^{49}$~ cm$^{-3}$, and a temperature,
    $T = 23$~MK, seen in Figure \ref{spectrum} (lower left panel).
    We calculated $V_{\rm th} = 6.14 \times 10^{26}$~cm$^{-3}$ as described
    in the previous paragraph to obtain a looptop density, $n_e = n_p= \sqrt{EM/V}= 4.42 \times
    10^{10}$~cm$^{-3}$. The spatial extent of the acceleration region was calculated
    to be $\sigma= 5.3 \times 10^8$~cm. These parameters are used as the
    input to our model corona.

  \subsection{Non-thermal spectral properties}\label{RHESSI_results_sect}

    \begin{figure*}[htpb]
    \centering
        \includegraphics[width=0.44\textwidth,clip=true, trim=0cm 0cm 0cm 0cm]{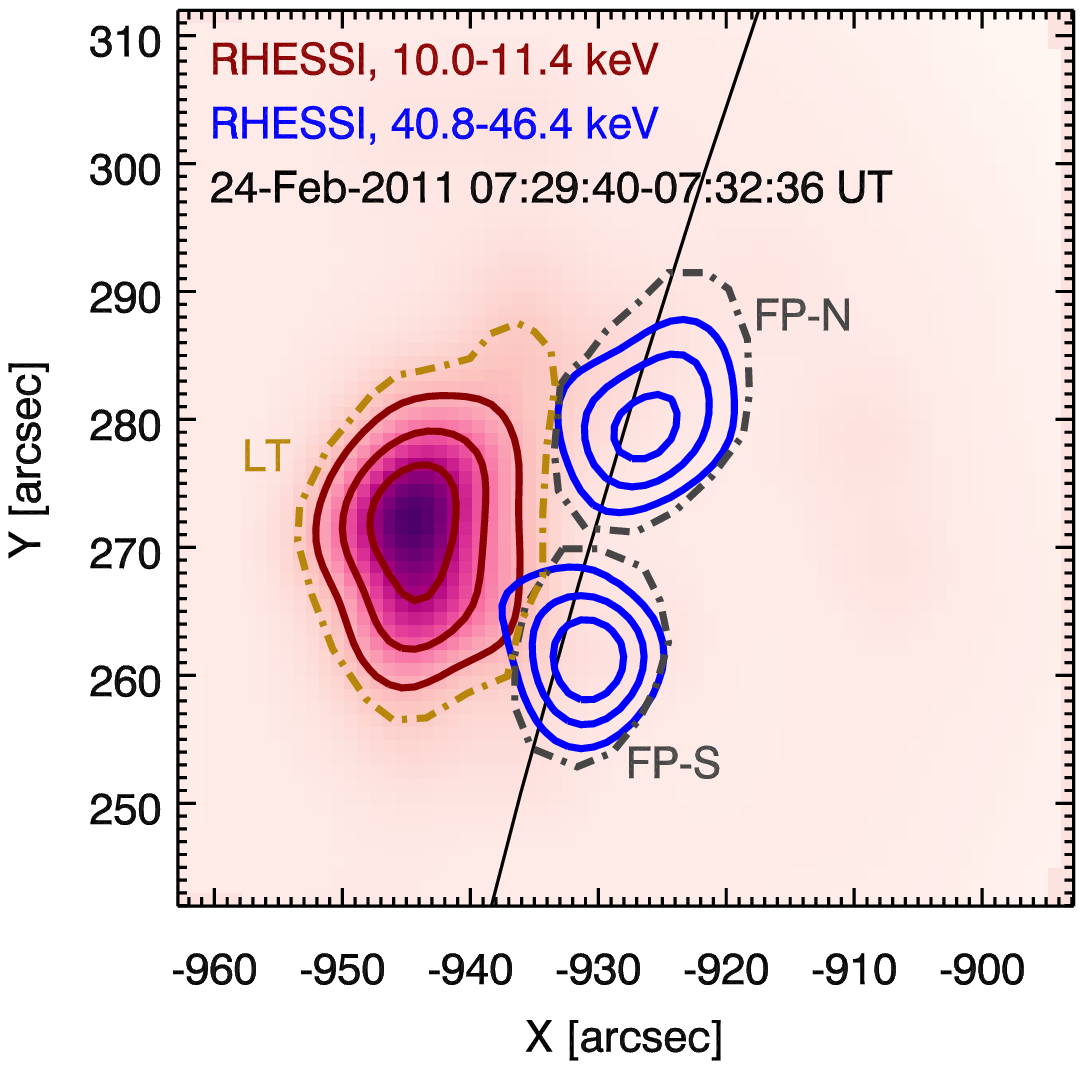}\\
        \includegraphics[width=0.32\textwidth,clip=true, trim=0cm 0cm 0cm 1cm]{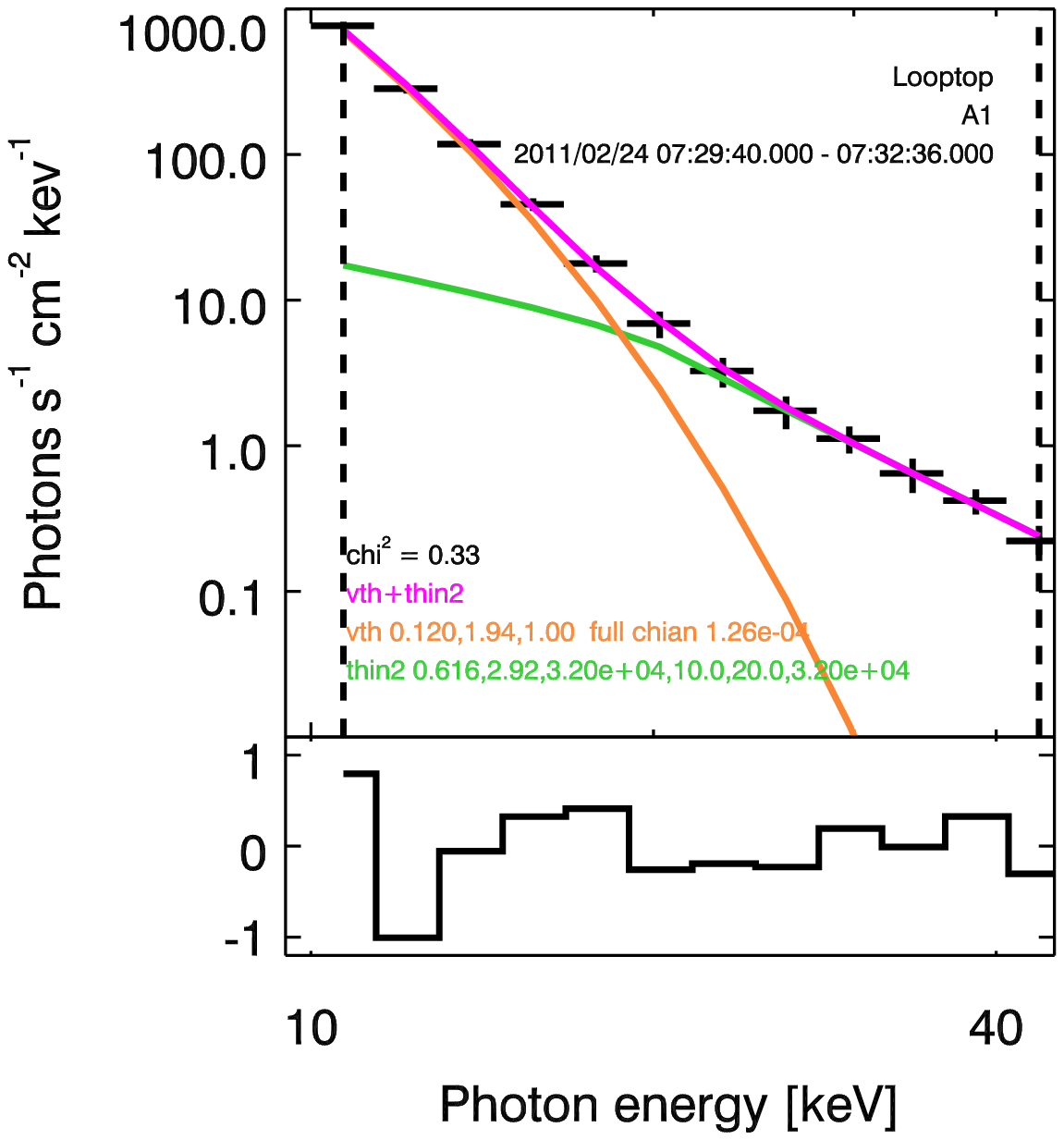}
        \includegraphics[width=0.33\textwidth,clip=true, trim=0cm 0cm 0cm 1cm ]{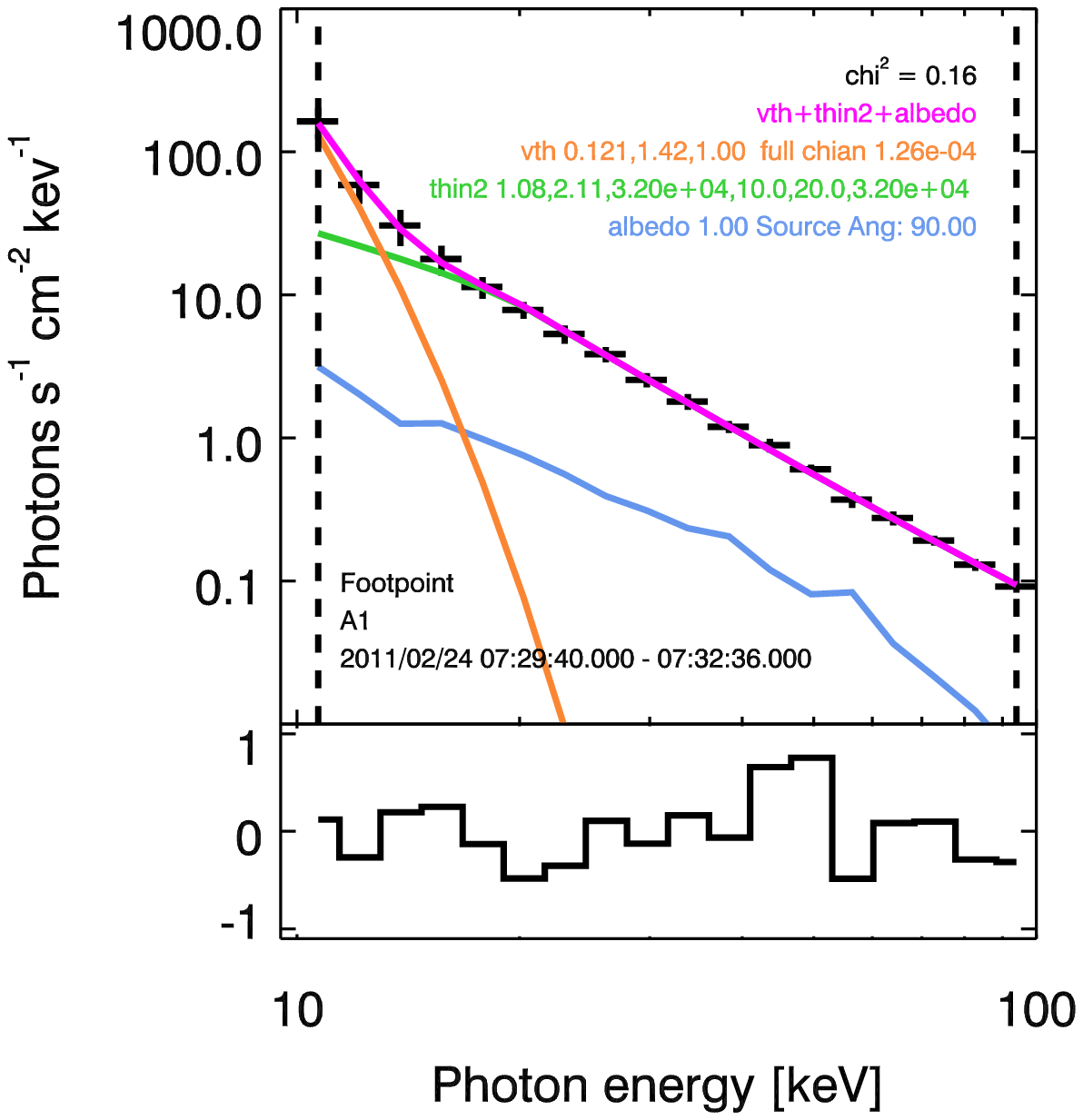}
        \includegraphics[width=0.33\textwidth,clip=true, trim=0cm 0cm 0cm 1cm]{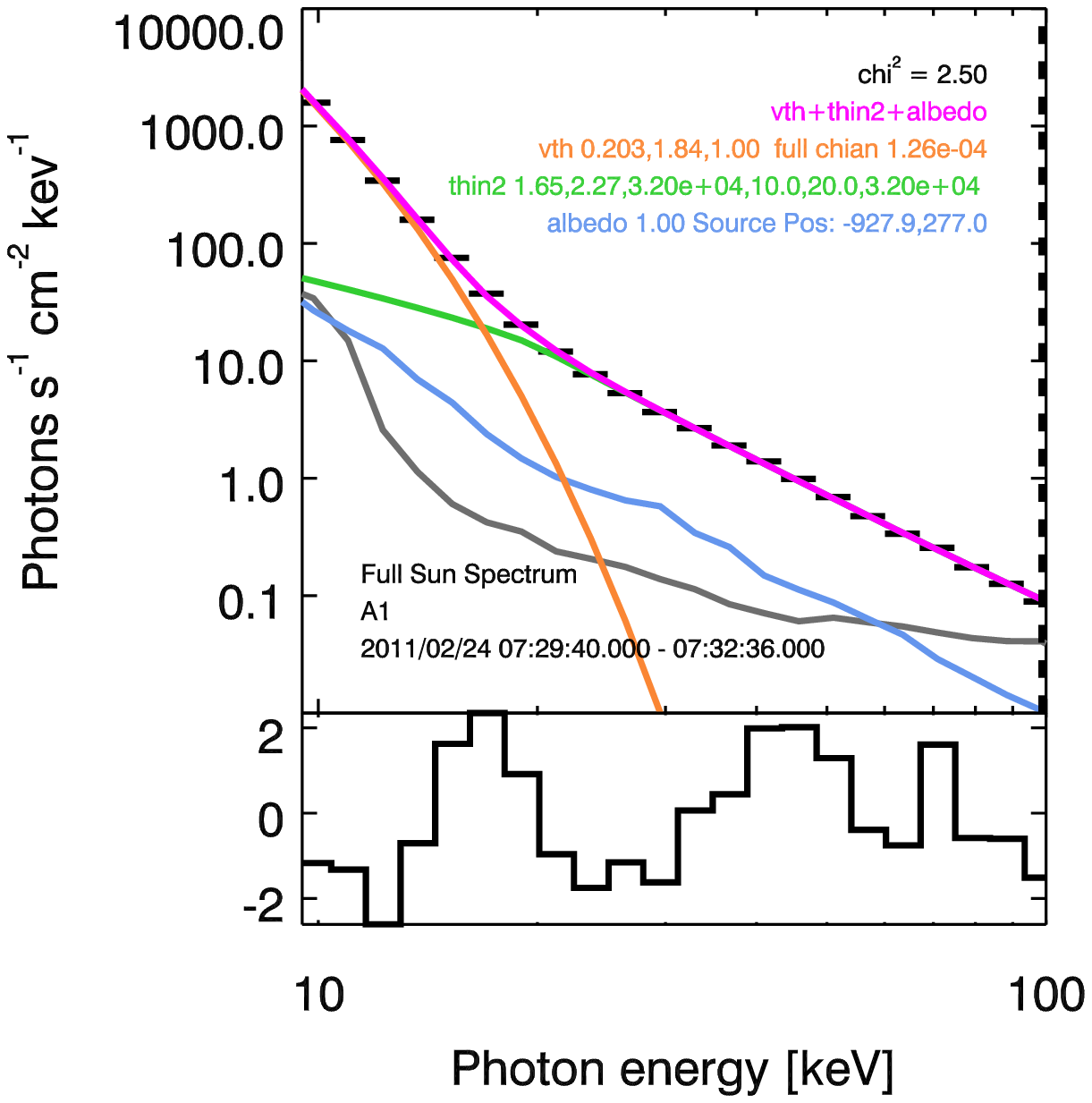}
        \caption{Top; CLEAN image of the $2011$ Feb $24$ flare. The dot-dash lines show the
        looptop (LT) and footpoint (FP-N, FP-S) regions used to produce spectra.
        The red contours show the looptop emission in the $10.0-11.4$~keV (30\%, 50\% and 75\%) energy
        band overplotted over the CLEAN image in the same range. The footpoint emission at
        $40.8-46.4$~keV is shown by the blue contours (30\%, 50\% and 75\%).
        Also shown are photon X-ray spectra for the flare: bottom left;
        looptop spectrum, bottom middle; summed footpoint spectrum and
        bottom right; full-Sun spectrum. HXR spectrum is shown as black data points.
        Fitting result is shown by the magenta line and is composed of a thermal (orange)
        and thin-target (green) component, with an albedo correction (blue) for the
        footpoint and full-Sun spectra. The full-Sun spectrum also shows the background emission in grey. The range fitted for each case is shown by the
        vertical dashed lines.}
        \label{spectrum}
    \end{figure*}

    The flux of non-thermal particles is
    $\langle n V F_0(E) \rangle_{\rm LT} = 0.62 \pm 0.15\times 10^{55}$~cm$^{-2}$ s$^{-1}$
    and the spectral index is $\delta_{\rm LT} = 2.91 \pm 0.43$.
    The footpoint sources, seen in the bottom middle panel in Figure \ref{spectrum},
    are best fit by a flux of $\langle n V F_0(E) \rangle_{\rm FP} = 1.08 \pm 0.06 \times 10^{55}$~cm$^{-2}$ s$^{-1}$
    and a spectral index of $\delta_{\rm FP} = 2.11 \pm 0.04$.
    The imaging spectroscopy
    results are consistent with the full-Sun spectrum seen in the bottom right panel in Figure \ref{spectrum}
    ($EM = (0.20 \pm 0.01) \times 10^{49}$~cm$^{-3}$, $T = 21$~MK,
    $\langle n V F_0(E) \rangle = (1.65 \pm 0.02) \times 10^{55}$~cm$^{-2}$ s$^{-1}$
    and $\delta = 2.27 \pm 0.01$).
    The low energy cutoff was fixed at $20$~keV for all fits.

    As expected the footpoint source has a harder spectrum
    of high energy electrons,
    but not by the factor two that would be expected if both
    coronal and footpoint photon spectra are fit with thin2 \citep[see][]{2013A&A...551A.135S}.
    This implies some kind of extra trapping within the coronal looptop
    source \citep{2013A&A...551A.135S,2013ApJ...777...33C}.
    Non-collisional pitch angle scattering in the presence of collisional
    losses hardens the electron spectrum in the coronal source at lower energies
    and this may result in both the
    looptop and footpoint spectra becoming broken power-laws \citep{1991ApJ...374..369B}.
    Fitting with a single power-law electron spectrum could thus result in spectral index
    differences between the looptop and footpoint sources that are not equal
    to two, as mentioned in \citet{2014ApJ...780..176K}. The spectra shown in
    Figure \ref{spectrum} show nothing in the residuals to suggest a break however,
    so it suffices to fit the non-thermal spectrum with a single power-law here.

    \section{Numerical solutions of the fokker-planck equation}\label{numerical_results}

\begin{figure}[htpb]
        \centering
        \includegraphics[width=0.49\textwidth, clip=true, trim=1.5cm 1.5cm 0cm 1.5cm]{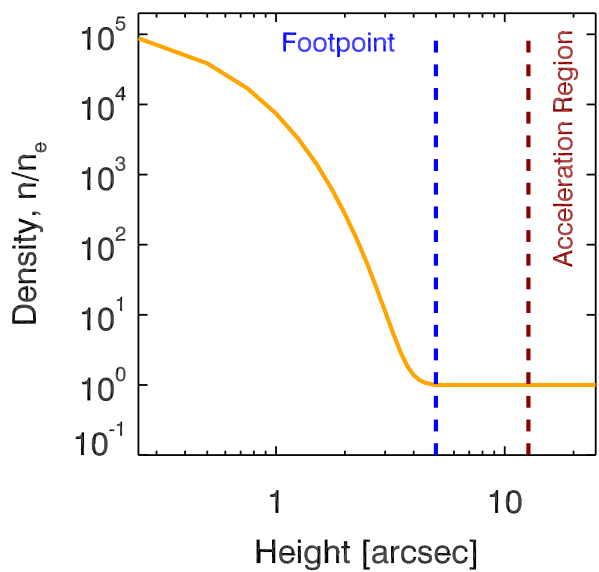}
        \caption{Density of the simulated corona, $n(x)/n_e$, as a function of height above the photosphere in arcseconds.
        Vertical lines show the boundaries of acceleration and footpoint regions.}
        \label{density_graph}
    \end{figure}

    We created a model corona with an originally Maxwellian distribution of
    particles at temperature, $T$. This means that the thermal speed $v_{\rm te}=\sqrt{k_{\rm B} T /m_{\rm e}}$ and,
    \begin{equation}\label{maxwellian}
    f = \sqrt{\frac{1}{2\pi v_{\rm te}^2}} \exp\left(-\frac{v^2}{2 v_{\rm te}^2} \right).
    \end{equation}
    The density, $n(x)$, is modelled as constant throughout the corona with an exponential increase at the chromosphere with
    scale height, $H = 220$~km, following a hydrostatic model consistent with RHESSI observations \citep{2012ApJ...760..142B},
    \begin{equation}
      n(x) = \begin{cases} n_e; & -5''\le x < 15'' \\  n_{\rm final} \exp\left(-\frac{|x-x_{\rm max}|}{H}\right)+n_e; &  15'' \le x \le 20'' \\  \end{cases},
    \end{equation}
    where $x_{\rm max}$ is the end of the numerical box ($20''$ in this case) and
    $n_{\rm final}$ is chosen to be sufficiently high to collisionally stop electrons.
    The density profile is shown in Figure \ref{density_graph}.
    The spatial extent
    of the acceleration region, $\sigma$, calculated above was used in Equation \eqref{dturb}.
    Setting $\alpha = 3$ for the reasons discussed in Section \ref{leaky_sect} we examined
    how the parameter $\tau_{\rm acc}$ affects the spectral index resulting from
    our simulations.
    The simulated index is compared to that predicted by the leaky-box
    solution (Equations \ref{scatt_free_LT_E_spect},
    \ref{isotropic_LT_E_spect}, \ref{scattfree_FP_E_spect} and
    \ref{isotropic_FP_E_spect}) valid for each transport
    regime to see how the introduction of a spatially inhomogeneous,
    extended acceleration region affects the distribution
    of the energized particles.
    The timescales shown here are $\tau_{\rm acc} =
    100, 180, 360, 900, 2000, 5000, 10000~ \tau^{\rm th}_{\rm c} $, where $\tau^{\rm th}_{\rm c} = v_{\rm te}^3/\Gamma$
    is the collisional timescale of a thermal electron in the corona,
    approximately $0.01$~s for the event
    in question ($\Gamma = 4 \pi e^4 \ln \Lambda n_e / m_{\rm e}^2$
    is the coronal collisional parameter here, independent of $x$).
    These timescales are chosen as they result in the range of electron spectral indices typically found in solar flares.
    The Fokker-Planck equations were solved numerically
    by the method of finite differences \citep{2001CoPhC.138..222K}.
    The results are discussed in Sections \ref{scatter-free_sect} and \ref{diffusive_sect},
    but first we discuss how to
    obtain $\langle n V F(E) \rangle$, and specifically the power law index,
    $\delta$, from the
    simulations to compare with the leaky-box solutions.

    The electron phase space distribution, $f(v,x)$, used in our simulations is
    directly related to the observed mean flux spectrum, so that
    the electron flux spectrum is $F(E)=f(v)/m_{\rm e}$ in
    the one-dimensional Fokker-Planck and $F(E)=v^2 f(v)/m_{\rm e}$ for the three-dimensional Fokker-Planck. The density weighted mean electron flux is,
        \begin{equation}
            \langle n V F(E) \rangle =  \int_V F(E,x) n(x)dV.
        \end{equation}
        So we have,
        \begin{equation}\label{coronal_source}
          \langle n V F(E)^{\rm CS}\rangle = A_{\rm LT} n_e \int_{-5''}^{15''} F(E,x) dx,
        \end{equation}
    where $A_{\rm LT}$ is the cross-sectional area of the loop and the limits are the estimation of the distance from the maximum emission in $10-11.4$~keV
    to one of the footpoints. The footpoint has a steeply increasing density
    (Figure \ref{density_graph}) over the last 5 arcseconds of our simulation
    domain.
    The density weighted mean electron flux from the model footpoint is thus,
        \begin{equation}\label{fp_source}
          \langle n V F(E)^{\rm FP}\rangle = A_{\rm LT}  \int_{15''}^{20''} F(E,x) n(x) dx.
        \end{equation}
    The power-law index of either the simulated looptop or footpoint source can then be found as,
        \begin{equation}
          \delta (E) = - \frac{ d \ln \langle n V F(E) \rangle}{ d \ln E},
        \end{equation}
    where the $E$ dependence of $\delta$ is to make clear that the simulated $\delta$ will
    not be constant with $E$ due to the extended, spatially varying nature of the acceleration region. The simulated spectral index was
    fit with a power-law between 25 and 50~keV, the reason being that the spectral index of a solar flare may be expected to vary with energy as well, so
    fitting our simulated $\langle n V F(E) \rangle$ with a power-law enabled a fairer comparison.
    The spectral index for each $\tau_{\rm acc}$ will be compared to the equivalent leaky-box solution with differences highlighted.
    In the next two sub-sections we summarise the simulation results for the different transport regimes.

    \subsection{Scatter-free transport}\label{scatter-free_sect}

    \begin{figure*}[htpb]
      \centering
        \includegraphics[width=0.49\textwidth, clip=true, trim=1.5cm 1.5cm 0cm 1.5cm]{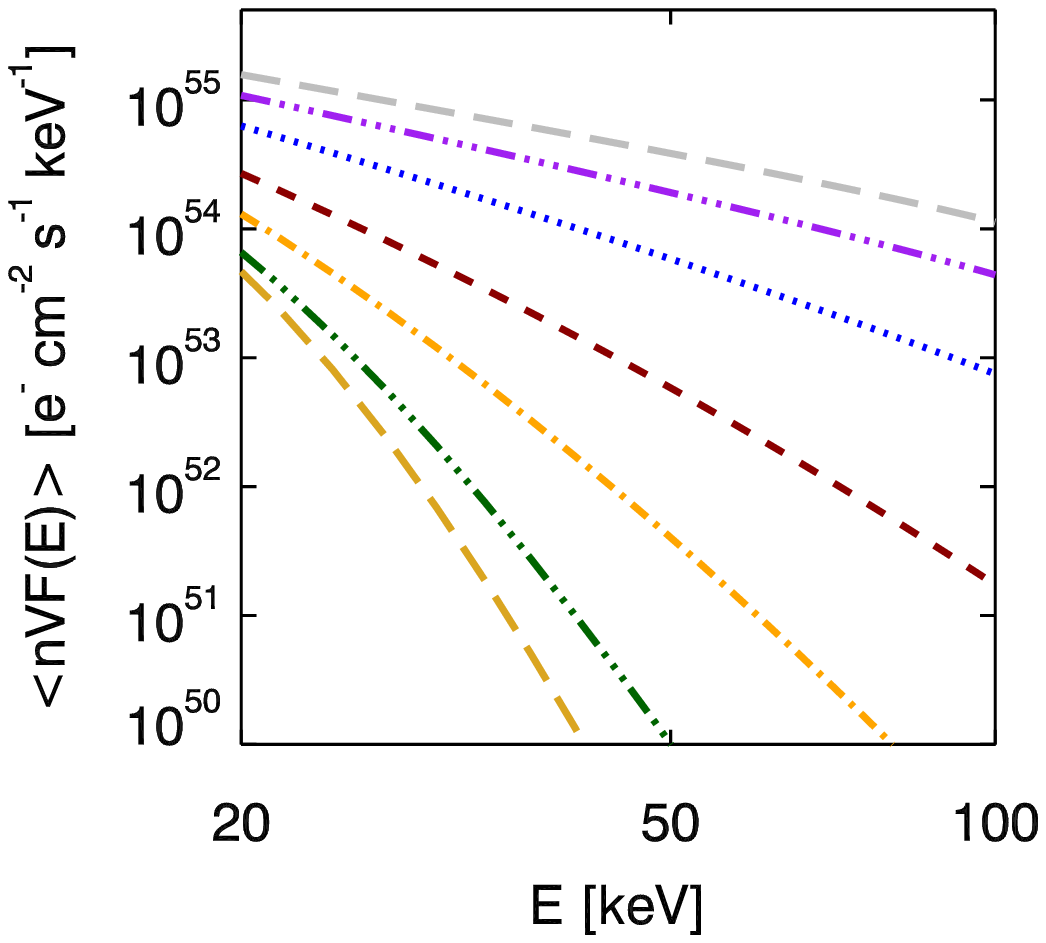}
        \includegraphics[width=0.49\textwidth, clip=true, trim=1.5cm 1.5cm 0cm 1.5cm]{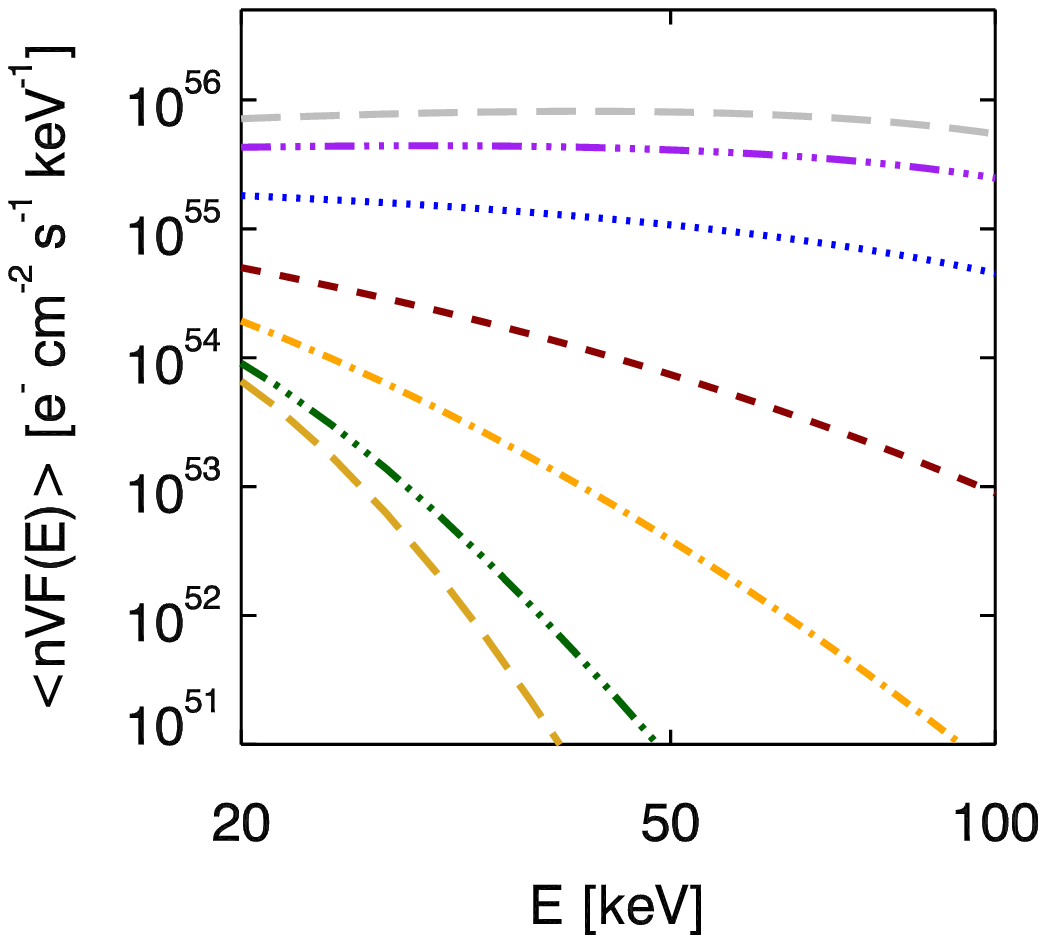}
        \caption{Simulated $\langle n V F(E) \rangle$ for the scatter-free
        transport case for coronal source (left) and footpoint
        source (right) for acceleration timescales: $\tau_{\rm acc} = 100 \tau^{\rm th}_{\rm c}$ (grey dash line),
        $\tau_{\rm acc} = 180 \tau^{\rm th}_{\rm c}$ (purple triple dot-dash line),
         $\tau_{\rm acc} = 360 \tau^{\rm th}_{\rm c}$
        (blue dot line), $\tau_{\rm acc} = 900 \tau^{\rm th}_{\rm c}$ (maroon dash line),
        $\tau_{\rm acc} = 2000 \tau^{\rm th}_{\rm c}$ (orange dot-dash line), $\tau_{\rm acc} = 5000 \tau^{\rm th}_{\rm c}$
        (green triple dot-dash line), $\tau_{\rm acc} = 10000 \tau^{\rm th}_{\rm c}$ (yellow long dash line).}
        \label{scatter-free_transport_fig}
    \end{figure*}

    \begin{figure*}[htpb]
      \centering
          \includegraphics[width=0.49\textwidth, clip=true, trim=1.5cm 1.5cm 0cm 1.5cm]{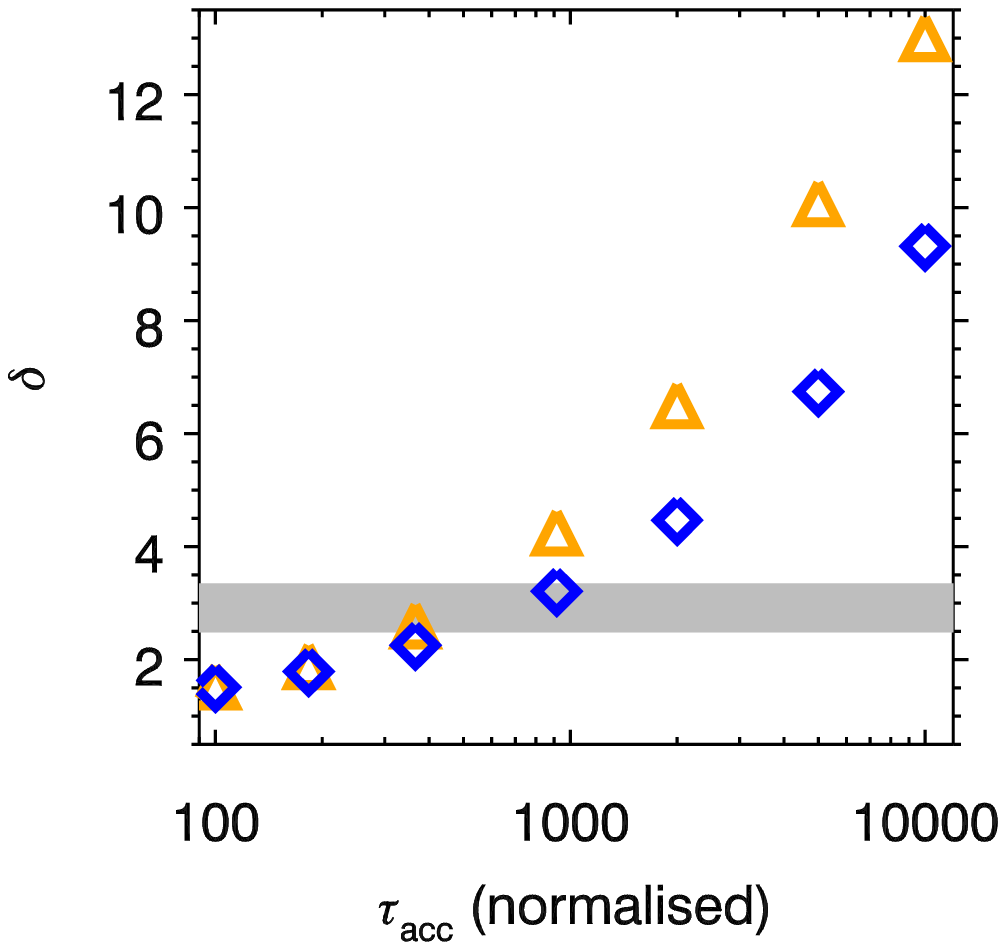}
          \includegraphics[width=0.49\textwidth, clip=true, trim=1.5cm 1.5cm 0cm 1.5cm]{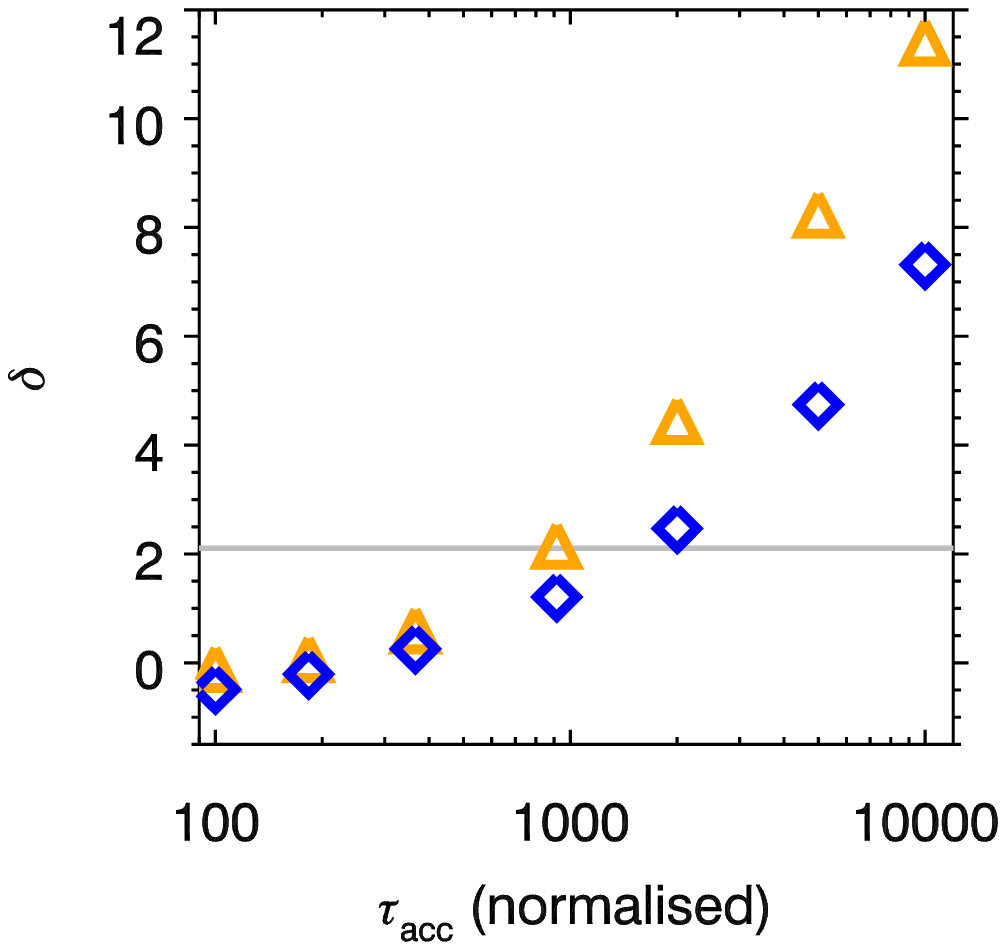}
        \caption{Spectral index, $\delta$, calculated from the mean electron flux shown in Figure \ref{scatter-free_transport_fig} for looptop (left) and footpoint (right).
        The grey confidence strip shows the
        possible range of $\delta^{\rm obs}$ for the fit ($\delta^{\rm obs}_{\rm LT} = 2.91 \pm 0.43$ and
        $\delta^{\rm obs}_{\rm FP} = 2.11 \pm 0.04$).
        Orange triangles are the $\delta$ obtained from fitting the simulated $\langle n V F(E) \rangle$ between 25 and 50~keV. The blue diamonds show the predicted spectral
        index from the leaky-box approximation.}
        \label{scatter-free_transport_spec_comp_fig}
    \end{figure*}

    Figure \ref{scatter-free_transport_fig} shows the simulated density weighted mean electron flux, $\langle n V F(E) \rangle$.
    This graph clearly illustrates the dependence of both the spectral index and non-thermal flux on the acceleration timescale, $\tau_{\rm acc}$.
    The longer the acceleration timescale, the less efficient the particle acceleration and the steeper the spectrum.

    The simulated spectral index from a power-law fit between 25 and 50~keV for each timescale studied is shown in Figure \ref{scatter-free_transport_spec_comp_fig} and is compared to the leaky-box
    solution (Equations \ref{scatt_free_LT_E_spect} and \ref{scattfree_FP_E_spect}). The fitted spectral index for the 2011 Feb 24 flare
    is overplotted for context.
    At short times, $\tau_{\rm acc} < 1000 \tau_{\rm c}^{\rm th}$, the spatially independent and inhomogeneous models agree for both the looptop and footpoint sources. However, there is a clear difference between the spatially independent leaky-box solution and the numerical solution to the
    Fokker-Planck (Equation \ref{1dfp}) at times $\tau_{\rm acc} \ge 1000 \tau_{\rm c}^{\rm th}$.
    For example, the fitted spectral index
    to the looptop is $\delta^{\rm obs}_{\rm LT} = 2.91 \pm 0.43$, the leaky-box Fokker-Planck predicts an
    acceleration timescale required to produce this spectral index of $\sim 1000 \tau_{\rm c}^{\rm th}$. That is to say
    this is the point where the blue diamonds overlap with the grey confidence band in Figure \ref{scatter-free_transport_spec_comp_fig} (left-panel). The numerical solution of the spatially
    inhomogeneous model (Equation \ref{1dfp}) predicts a softer index here and, as such, this model would
    require a shorter acceleration timescale, somewhere between $300< \tau_{\rm acc} < 1000 \tau_{\rm c}^{\rm th}$
    to produce the observed looptop
    spectral index. This is particularly pertinent when one considers that typical looptop spectra are, in
    general, softer than that observed here, the typical range being $\delta^{\rm obs}_{\rm LT} \sim 2-8$ \citep[see e.g.][]{2006A&A...456..751B}. Therefore,
    in the range of `realistic' spectral indices the spatially inhomogenous model produces
    softer spectra than the spatially independent model. Furthermore, the same behaviour is seen in the right
    panel of Figure \ref{scatter-free_transport_spec_comp_fig} for the footpoint spectrum, where it is clear
    that for the observed $\delta^{\rm obs}_{\rm FP}$ the spatially independent model predicts a longer
    acceleration timescale than our spatially inhomogeneous model. It is also important to note that footpoint spectral indices $\ll 2$
    are rare; in the right panel of Figure \ref{scatter-free_transport_spec_comp_fig} it is easy to see that in the range $\delta_{\rm FP} \ge 2$
    there is a substantial difference between the acceleration timescales required by both models to produce the same spectral index.

    In summary, the introduction of spatially inhomogeneous acceleration and transport
    reduces the acceleration efficiency compared to the spatially independent leaky-box formulation for the
    standard range of spectral indices observed by RHESSI.
    As a result, any acceleration timescale inferred from the
    the leaky-box approximation could be an overestimate of the actual acceleration timescale in the flare.

    \subsection{Diffusive transport with $\lambda = 5 \times 10^8$~cm}\label{diffusive_sect}

    \begin{figure*}[htpb]
      \centering
      \includegraphics[width=0.49\textwidth, clip=true, trim=1.5cm 1.5cm 0cm 1.5cm]{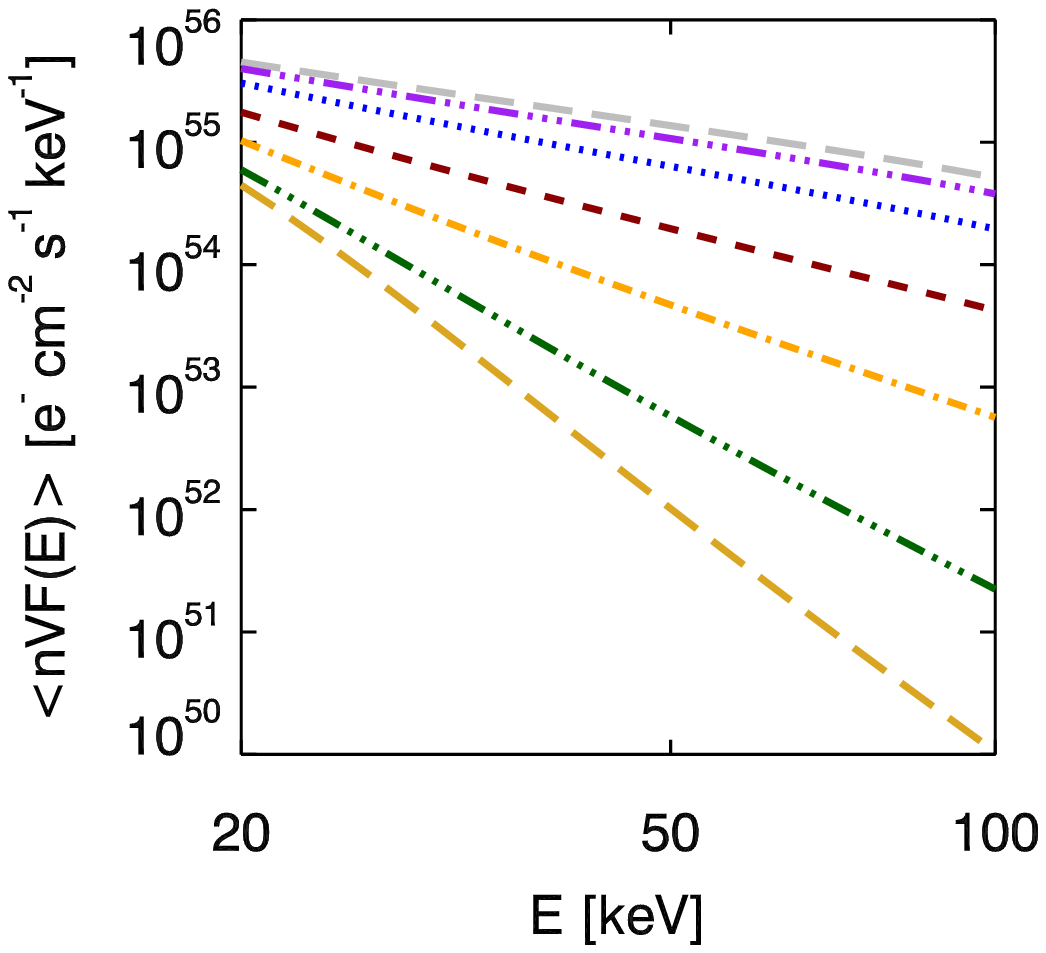}
      \includegraphics[width=0.49\textwidth, clip=true, trim=1.5cm 1.5cm 0cm 1.5cm]{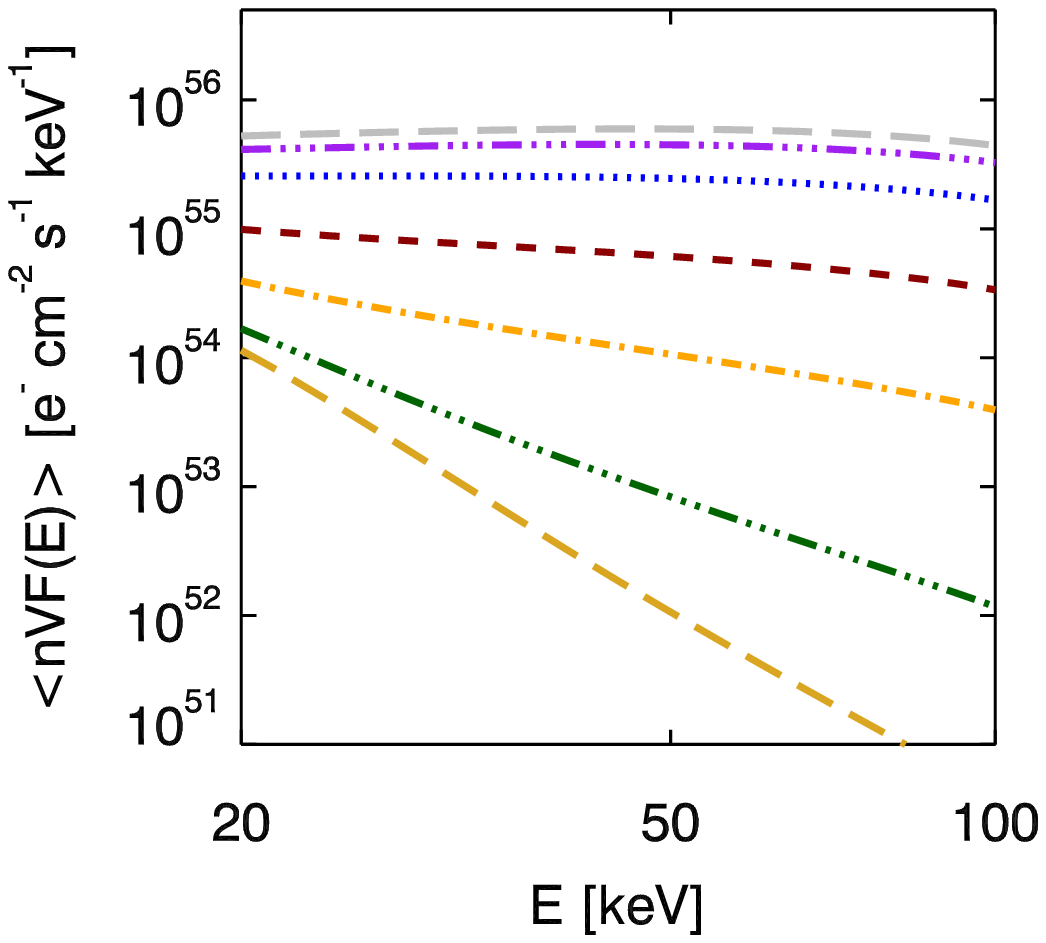}
      \caption{Same as Figure \ref{scatter-free_transport_fig} but for diffusive transport
      with constant mean free path in velocity, $\lambda = 5 \times 10^{8}$~cm.}
      \label{diffusive_transport_fig}
    \end{figure*}

    \begin{figure*}[htpb]
      \centering
      \includegraphics[width=0.49\textwidth, clip=true, trim=1.5cm 1.5cm 0cm 1.5cm]{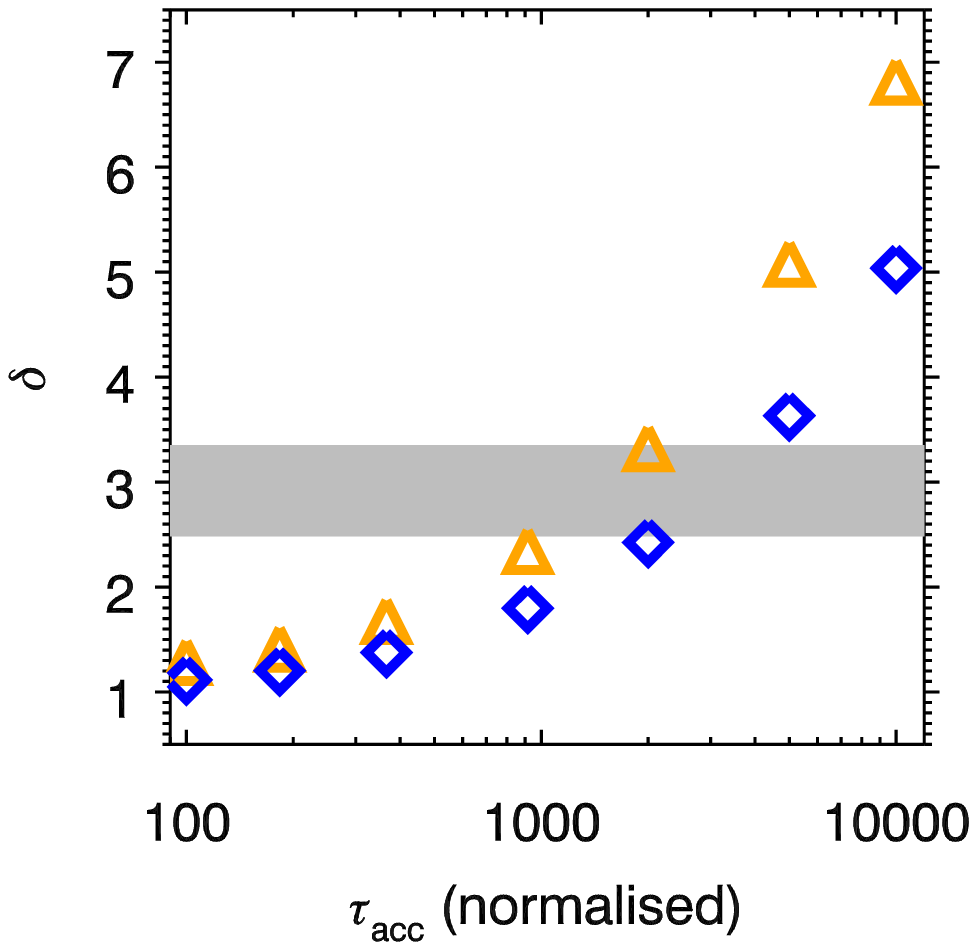}
      \includegraphics[width=0.49\textwidth, clip=true, trim=1.5cm 1.5cm 0cm 1.5cm]{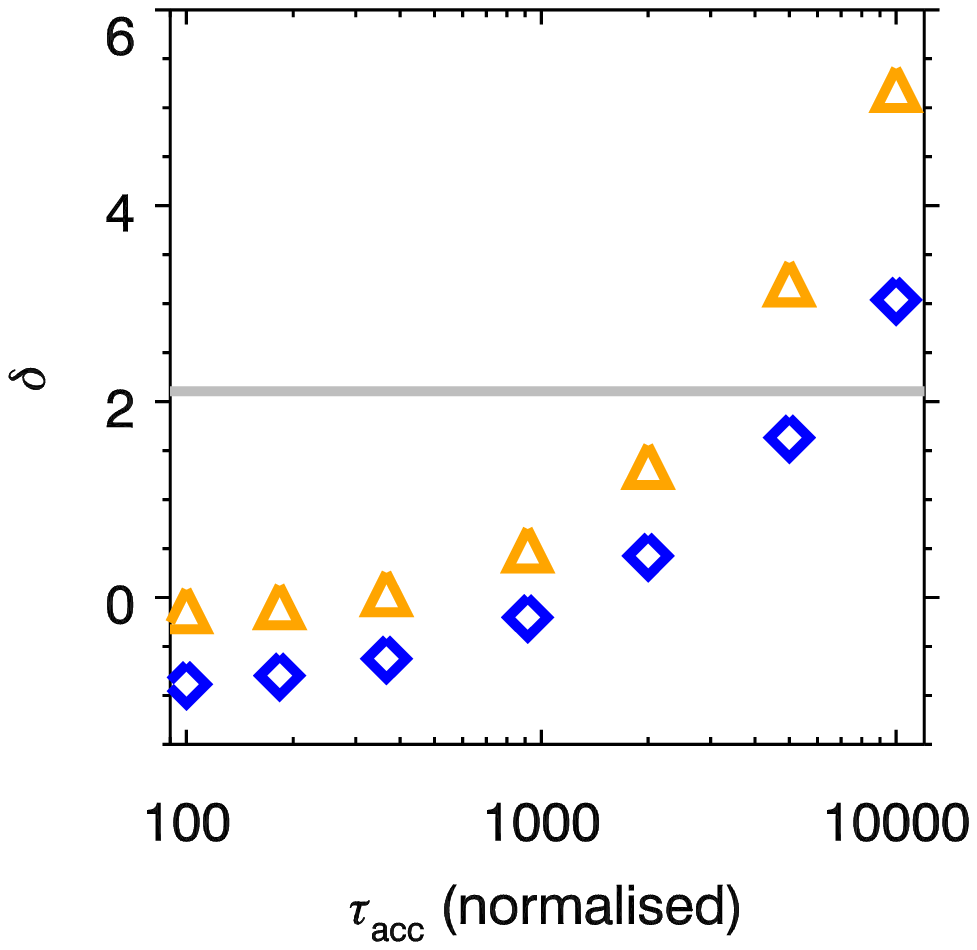}
      \caption{Same as Figure \ref{scatter-free_transport_spec_comp_fig} but for diffusive
      transport with constant mean free path in velocity, $\lambda = 5 \times 10^{8}$~cm.}
      \label{diffusive_transport_spec_comp_fig}
    \end{figure*}

    Figure \ref{diffusive_transport_fig} shows the $\langle n V F(E) \rangle$ for diffusive transport
    with a constant mean free path of $\lambda = 5 \times 10^8$~cm. This value is chosen due \citet{2014ApJ...780..176K} finding it the midpoint of the limits for 30~keV electrons, as discussed earlier. Like Figure \ref{scatter-free_transport_fig} we clearly see the relationship between the flux of non-thermal particles, the spectral index and the acceleration
    timescale.

    The simulated spectral index is compared to that predicted from the leaky-box Fokker-Planck approximation
    (Equations \ref{isotropic_LT_E_spect} and \ref{isotropic_FP_E_spect}) and the imaging
    spectroscopy results from the 2011 February 24 flare in
    Figure \ref{diffusive_transport_spec_comp_fig}. We again see
    a similar behaviour to the scatter-free case for both the spatially independent
    and inhomogeneous models with respect to the acceleration timescale.
    For the realistic range of $\delta^{\rm obs}$ discussed in the previous section there is again a large
    difference between the index predicted by the leaky-box analytic solution and the spatially inhomogeneous
    model considered (Equation \ref{3dfp}). Specifically, when spatial effects are fully taken into account
    the model produces a generally softer spectral index than the spatially independent leaky-box formalism. Thus, we again conclude that a
    timescale inferred when using the leaky-box model could in fact be an overestimation of the actual
    acceleration timescale in the system.

    It is important to note that the (slightly) negative spectral indices obtained for the numerical simulations in Figures \ref{scatter-free_transport_spec_comp_fig} and \ref{diffusive_transport_spec_comp_fig} arise from the shortest
    acceleration timescales studied (see grey and purple lines in Figures \ref{scatter-free_transport_fig} and \ref{diffusive_transport_fig}). Such spectral indices are not observed, however, and so the respective timescales are too short for the solar flare case.

 \section{Summary}\label{conc_sect}

    In this paper we introduced a model accounting for the intrinsic spatial variation in the acceleration region of solar flares. By using the imaging spectroscopy
    of the 2011 February 24 flare the density, temperature and spatial extent of the acceleration region were inferred and used as input parameters to the model.
    We solved the governing kinetic equations numerically,
    and compared to the spatially invariant leaky-box approximation
    commonly used when studying stochastic acceleration in solar flares. The results are summarised as follows:

    \begin{itemize}
      \item Scatter-free transport; the introduction of a spatially inhomogeneous acceleration region
      while explicitly accounting for transport results in acceleration that is generally less efficient than the spatially independent leaky-box formulation. The resulting spectral index, for both looptop and footpoint sources, is softer than that when spatial effects are not explicitly taken into account.

      \item Diffusive transport with $\lambda=5 \times 10^8$~cm; similar behaviour is seen for the diffusive transport case, the introduction of a spatially extended, inhomogeneous, acceleration region results in a spectrum that is softer, for the most part, than that predicted by the leaky-box solution.
    \end{itemize}

    In summary, for both transport regimes studied it is clear that the intrinsic spatial dependency evident in
    solar flares \citep{2008ApJ...673..576X,2012A&A...543A..53G} changes the resulting electron spectrum when compared to the spatially independent leaky-box approximation. It acts to reduce the
    acceleration efficiency and thus produces a softer spectrum. This is particularly pronounced in the `standard' range of spectral indices, $\delta$, generally observed by RHESSI ($2 \lesssim \delta^{\rm obs}_{\rm LT} \lesssim 8$ and $\delta_{\rm FP}^{\rm obs} \gtrsim 2$, see e.g. \citealp{2006A&A...456..751B}). This means that the acceleration timescales inferred when using a leaky-box model applied to a solar flare could be an overestimation.
    These timescales should therefore be considered an upper limit of the time taken
    to produce the observed spectral index. Thus, the authors suggest that the intrinsic
    spatial dependence should be taken into account when modelling stochastic acceleration in solar flares.

\begin{acknowledgements}
 This work was supported by the STFC, via an STFC consolidated grant (EPK) and an STFC studentship (DJS). The work was undertaken at the University of Glasgow, and for that the authors express gratitude.
 DJS further wants to
 thank Researchers in Schools and King's College London for honourary visiting researcher status, as well as the Co-op Academy Failsworth for allowing the continuation of research in a new career. Both authors would also like to thank the referee for their helpful comments to improve the paper.
\end{acknowledgements}

\bibliographystyle{aa}
\bibliography{ref_paper}

\end{document}